\documentclass[aps,amsmath,amssymb,superscriptaddress,twocolumn,nofootinbib,showkeys,showpacs,floatfix]{revtex4}

\usepackage{graphicx}
\usepackage{textcomp}
\usepackage{times}
\usepackage{hyperref}

\begin{document}

\title{Determination and prediction of the fretting crack initiation: introduction of the (P,Q,N) representation and definition of a variable process volume}

\author{H. Proudhon}
\email{henry.proudhon@ec-lyon.fr}
\affiliation{Laboratoire de Tribologie et de dynamique des syst\`{e}mes, UMR 5513, Ecole Centrale de Lyon, 36~Avenue Guy de Collongue, 69134 Ecully Cedex, France}

\author{S. Fouvry}
\thanks{Corresponding author. Tel.: +33-472-186-562; fax: +33-472-433-383.}
\affiliation{Laboratoire de Tribologie et de dynamique des syst\`{e}mes, UMR 5513, Ecole Centrale de Lyon, 36~Avenue Guy de Collongue, 69134 Ecully Cedex, France}

\author{G. R. Yantio}
\affiliation{Laboratoire de Tribologie et de dynamique des syst\`{e}mes, UMR 5513, Ecole Centrale de Lyon, 36~Avenue Guy de Collongue, 69134 Ecully Cedex, France}

\begin{abstract}
In this work, the crack nucleation under fretting loading is investigated experimentally with a damage tolerant 2024 aluminium alloy. A new method is introduced to determine its condition with respect to all loading parameters including the number of fretting cycles. Further work deals with the prediction of this threshold using the Smith-Watson-Topper criterion. New developments are presented, in particular a process volume of variable size is introduced in the computations of the fretting crack initiation.
\end{abstract}

\maketitle

\section{Introduction}\label{intro.sec}

Fretting damage has been recognized as a problem in several industrial applications for years now. The first studies defined the concept of fretting \cite{Waterhouse1981}, while further work introduced more rationalization with precise concepts such as fretting cycles and fretting maps \cite{Vincent1992,Fouvry1996}. Depending on the contact parameters and especially the displacement amplitude, it has been shown that two main regimes can be defined. First, In the large displacement range, the full sliding regime exists, which can induce wear of the surfaces in contact. Second, In the low displacement amplitude range, the contact causes a partial slip condition and severe stress gradients at the contact border. This can induce very rapid crack nucleation compared to classical fatigue testing. In this paper we focus on the latter regime and particularly on the crack nucleation condition. This has been recognized as a critical issue for industry in the past ten years.

Thanks to recent work, different methods are now available to model the propagation of fretting fatigue cracks using for example the ``\emph{Crack analogue}'' model, fracture mechanics \cite{Giannakopoulos1998,Ciavarella2001} and also weight functions \cite{Navarro2003}. But due to the possible drastic reduction of the fatigue limit induced by fretting \cite{Lindley1992}, a great part of the work has been devoted to determining the fretting crack initiation conditions in various types of materials \cite{Szolwinski1996,Fouvry1998,Bernardo2004}. This process is quite complex due to high stress and strain gradients, oxidation phenomena and other tribological phenomenon (TTS formation \cite{Sauger2000}), surface deformation and wear, and debris formation. Until now the main method used to describe quantitatively crack nucleation deals only with the mechanical state under the fretting conditions and relies on computing a multiaxial fatigue criterion (this was first attempted by Petiot et al. \cite{Petiot1995}). This can be done either by using analytical solutions of the stress/strain state under fretting contact (as it can be found in \cite{Hills1994}), or more recently by numerous author with finite elements \cite{Naboulsi2003,Bernardo2004,Sum2005}. The latter method is more constrained but allows one to test more general geometries and more complex loading conditions. Due to the very strong gradients located at the contact border, one must use a process volume approach, which consists of averaging the stress/strain state in a micro-volume of material prior applying the multiaxial criterion. This method has popularized and has been investigated with various fatigue criteria \cite{Fouvry1998,Naboulsi2003,Proudhon2005}. These investigations have been capable of predicting, in some cases, the crack nucleation threshold~; but, the significance of the process volume is not yet fully understood, and as a consequence cannot be truly estimated without doing experiments. For completeness, it must be pointed out that some alternative methods exists such as the asymptotic stress intensity analysis \cite{Mugadu2002}. In this kind of approach, the stress singularity induced by the contact configuration is captured by asymptotic analysis in order to deduce a stress intensity factor at the contact edge.

In this paper we restrict the analysis to the prediction of the crack nucleation condition in the  fretting wear configuration by the Smith-Watson-Topper criterion. We present an new vision of the process volume and apply the method to a complete experiment series with various fretting wear conditions. This work attempts to clarify the physical meaning of the process volume and to set up a systematical way of characterizing the crack nucleation condition with respect to parameters relevant for the industrial applications (contact pressure, shear stress amplitude and number of cycles).

\section{Experimental work}
\label{exp_N.sec}

\begin{figure}[hbt]
\centering
\includegraphics{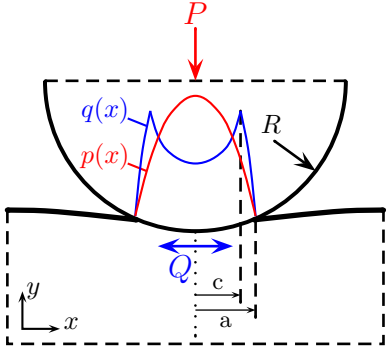}
\caption{Schematic of the fretting wear cylinder/plane configuration.}
\label{fretting_schema_stress.fig}
\end{figure}

Part of the experimental work has been presented elsewhere. The reader is referred to \cite{Proudhon2005} for more details on the experimental setup, though general information is summarized here. A general fretting wear apparatus is used in cylinder/plane configuration and under partial slip regime. A normal force P is applied on the counter-body to maintain the surface in contact causing an elliptic pressure $p(x)$ over the contact zone $|x|<a$~; a cyclic relative displacement is imposed to the interface leading to a classical shear stress $q(x)$ which exhibits its maximum at the stick zone limit $x=\pm c$ (cf. fig.~\ref{fretting_schema_stress.fig}). The flat specimen being maintained in a fixed position, this displacement gives rise to a cyclic shear force\footnote{note that in fretting wear experiments, there is no bulk load imposed} of magnitude $Q$ which is measured during the test via a force sensor. Regarding the materials, an aluminium/aluminium contact is studied, with plane samples in a 2024T351 alloy (see table \ref{alu_mecha.tab} for mechanical properties) and a cylindrical counter body of 49 mm radius made of a Al7075T6 alloy.

\begin{table}[htb]
\caption{Mechanical properties of the studied Al2024}
\begin{tabular}{cccc}
\hline
$E$ (GPa) & $\nu$ & $\sigma_{0.2\%}$ (MPa) & $\sigma_d$ (MPa) \\
\hline
72 & 0.33 & 325 & 140 \\
\hline
\end{tabular}
\label{alu_mecha.tab}
\end{table}

\begin{figure}[hbt]
\centering
\includegraphics[width=75mm]{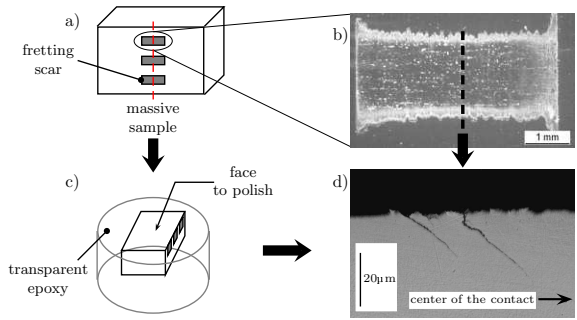}
\caption{Experimental method to investigated cracking after a fretting test~; see the text for details.}
\label{fretting_damage.fig}
\end{figure}

The fretting crack nucleation has been investigated in the following way (cf. fig.~\ref{fretting_damage.fig}):
\begin{itemize}
\item fretting test is conducted on the sample for chosen loading conditions (fig.~\ref{fretting_damage.fig}a).
\item fretting crack presence is investigated by cutting the sample in the middle of the scars and polishing the newly created face (fig. \ref{fretting_damage.fig}b,c).
\item chemical etching can be performed after a first optical micrograph observation in order to avoid a possible blurring of the crack by the polishing process (fig. \ref{fretting_damage.fig}d).
\item depending on the previous investigations, new fretting conditions are determined to refine or confirm the results.
\end{itemize}

A previous study (with the same geometry and material) conducted to 50\,000 fretting cycles was described in \cite{Proudhon2005}. According to these experiments, the crack nucleation condition was found to be mainly independent of the normal load (see fig.~\ref{prediction_r_variable.fig} where the experimental results have been replotted), leading to the crack nucleation threshold in terms of the tangential load $Q_{th}\simeq 240$ N/mm.

Though this value defines a boundary between safe and a crack risk zones, the number of cycle needed to be more deeply investigated. Indeed a rapid calculation of the stress state at the contact border does show that the contact is subjected to loading conditions equivalent to low cycle fatigue. We consider the simple analysis of the two-dimensional plane strain cylinder/flat contact with the Hertz theory from which we can write the peak contact pressure $p_0$:

\begin{equation}
p_0=\frac{2P}{\pi a}=\left(\frac{PE^*}{\pi R}\right)^{1/2}
\label{po.eqn}
\end{equation}

$R$ being the radius of the cylinder and $E^*$ the effective Young modulus.

The stress field is known in partial slip conditions from the analytical solutions (see for instance \cite{Hills1994}). At the contact border ($x=a,y=0$), $\sigma_{yy}=\sigma_{xy}=0$ so the stress tensor matches the principal stress state ($\sigma_1,\sigma_2,\sigma_3$) which is biaxial (the contact pressure vanishes):

\begin{align}
\label{biaxial.eqn}
\sigma_1 & =\sigma_{xx}=2p_0\sqrt{\frac{\mu Q}{P}}  \nonumber\\
\sigma_2 & =\sigma_{yy}=0 \\
\sigma_3 & =\sigma_{zz}=\nu\sigma_{xx} \nonumber
\end{align}

with $\mu$ the friction coefficient in partial slip condition ($\mu$ was determined to be close to 1,1 in our case \cite{Proudhon2005}). The equivalent Von Mises stress $\sigma_e$ can be easily expressed:

\begin{align}
\sigma_e & =\frac1{\sqrt{2}}\left((\sigma_1-\sigma_3)^2+\sigma_1^2+\sigma_3^2\right)^{1/2} \nonumber\\
         & =\frac1{\sqrt{2}}\left(2\sigma_1^2+2\sigma_3^2-2\sigma_1\sigma_3\right)^{1/2} \nonumber\\
         & = \sigma_1\,\left(1+(\frac{\sigma_3}{\sigma_1})^2-\frac{\sigma_3}{\sigma_1}\right)^{1/2}
\label{VM_expr.eqn}
\end{align}

combining with equation \ref{biaxial.eqn}:

\begin{align}
\sigma_e(x=a,y=0) & = 2p_0\sqrt{\frac{\mu Q}{P}}\,(\nu^2-\nu+1)^{1/2} \nonumber \\
                  & = 2p_0\sqrt{\frac{\mu E^* Q}{\pi R}}\,(\nu^2-\nu+1)^{1/2}
\label{VM_max.eqn}
\end{align}

Thus the shear force leading to the conventional yield at 0.2\% of deformation $\sigma_Y$, in the slip zone, can be written as:

\begin{equation}
Q_Y=\frac{\pi R}{4\mu E^*}\frac{\sigma_Y^2}{\nu^2-\nu+1}
\label{pression_plast.eqn}
\end{equation}

With $\sigma_Y=\sigma_{0.2\%}$, this expression gives $Q_Y=117$ N/mm. It is clear that an aluminium/aluminium contact with a friction coefficient greater than unity (like in our case) will induce very high contact stresses and it can be seen that all the tests carried out in this study operates under low cycle fatigue range ($Q>Q_Y$). It is therefore expected that the number of fretting cycles must have an influence on the crack nucleation condition.

\begin{figure}[hbt]
\centering
\includegraphics[width=75mm]{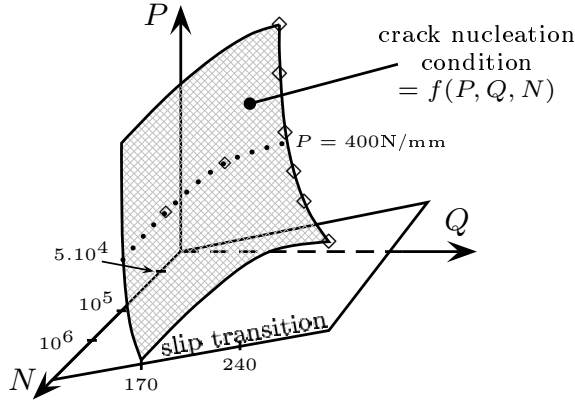}
\caption{Methodology used to determine the crack nucleation condition with respect to the $,Q,N$ parameters, a) tests at constant N, b) tests at constant P, c) whole (P,Q,N) representation.}
\label{fretting_nucleation.fig}
\end{figure}

To fully determine the crack nucleation condition, the experimental study was completed as follows: for a chosen normal pressure, additional fretting tests were conducted with different numbers of cycles. The initiation threshold is then determined for each new number of fretting cycles. The crack nucleation boundary is then expressed as a function of the three loading parameters $P$, $Q$ and $N$.
The whole experimental procedure is described in fig.~\ref{fretting_nucleation.fig}.

\begin{figure}[hbt]
\centering
\includegraphics{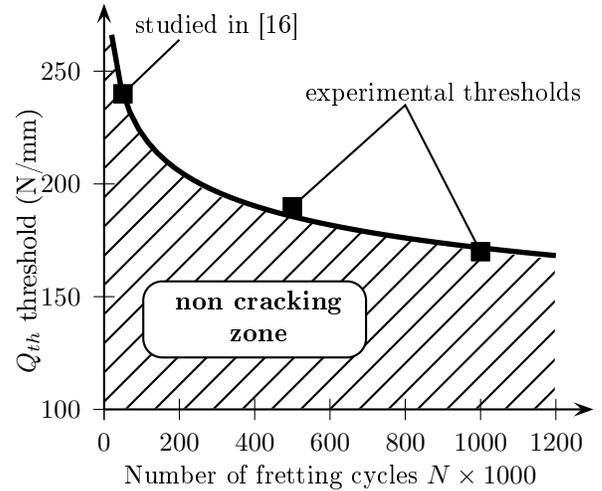}
\caption{Evolution of the critical tangential force $Q_{th}$ to initiate a fretting crack (determined as shown in fig.~\ref{fretting_nucleation.fig}a), with the number of fretting cycles ($P=400$ N/mm).}
\label{fretting_influence_N.fig}
\end{figure}

Here, new tests have been performed with a constant normal load $P=400$ N/mm. Two different numbers of fretting cycles have been investigated: $5.10^5$ and $10^6$ cycles. The corresponding crack nucleation threshold are determined according to fig.~\ref{fretting_nucleation.fig}a~; values have been reported in table \ref{fretting_amorcage_N.tab} and plotted on fig.~\ref{fretting_influence_N.fig}.

\linespread{1}
\begin{table}[htb]
\caption{Influence of the number of fretting cycles on the crack nucleation threshold ($P=400$ N/mm).}
\begin{tabular}{ccc}
\hline
Number of cycles & Threshold (N/mm)\\
\hline
$5.10^4$ & $240$ \\
$5.10^5$ & $190$ \\
$1.10^6$ & $170$ \\
\hline
\end{tabular}
\label{fretting_amorcage_N.tab}
\end{table}

As expected, the number of applied fretting cycles has an influence on $Q_{th}$. The more fretting cycles applied, the less the threshold value. Moreover, the evolution of $Q_{th}(N)$ shows first a rapid reduction with increasing $N$, and then tends to a saturation value around 160--170 N/mm ,although more experiment are needed to establish the exact value. In particular, experiments with $N>10^6$ cycles, which are very time consuming, are required. In addition, it has been shown elsewhere that with this material in this configuration, $10^6$ cycle does correspond to a stabilized condition \cite{Munoz2006}.

\begin{figure}[hbt]
\centering
\includegraphics[width=75mm]{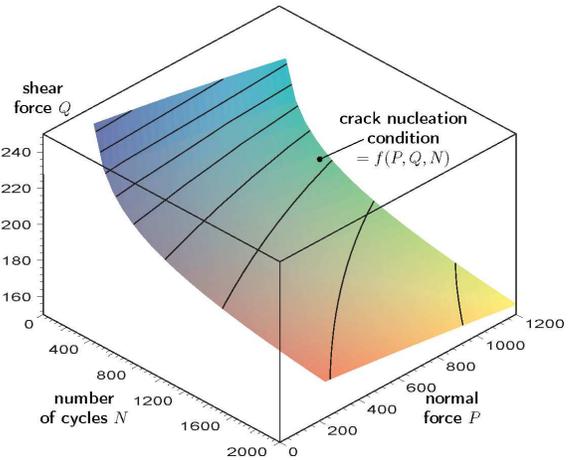}
\caption{Representation of the crack nucleation condition in the (P,Q,N) space.}
\label{fretting_PQN.fig}
\end{figure}

Assuming a homotetic behaviour with respect to $P$, the whole crack nucleation condition can be plotted in a 3D diagram $f(P,Q,N)$ as shown in figure~\ref{fretting_PQN.fig}.

This defines completely both the safe and cracking domains by the boundary $f(P,Q,N)$. The advantage is the strong physical meaning of these parameters, which can be extended to other test configurations and more generally to any structure. In fretting fatigue tests for instance, the definition of $P$ is unchanged but there is an additional load (bulk stress) in the material; the parameter $Q$ must take that bulk stress into account and $N$ would be the number of fatigue cycles. In a more complicated case like an industrial component, first step will be to locate the contact area and then calculate the contact pressure $P$, for example by finite elements (integrating the normal stresses over the contact surface). Then $Q$ can be defined as the maximum shear force transmitted through the contact during a cycle and should be estimated in the same way, integrating this time the shear stresses over the contact surface (this will, however, require a detailed contact analysis which must capture the partial slip contact conditions). $N$ can then be identified as the number of cycles associated to the load responsible of the relative displacement of the surfaces. To conclude, this new representation aims to narrow the gap between ideal test configuration and actual structures encountered in real industrial problems.

\section{Prediction of the crack nucleation condition}

The purpose of this section is to use a multiaxial fatigue criterion to predict the crack nucleation condition. The stress/strain state is computed from the analytical solutions of the cylinder plane contact~; these fields are averaged\footnote{this process is sometimes referred to as a size effect} on a circular shaped process volume to soften the contact gradients and then the cracking risk is evaluated. Among the numerous studies on the subject, it has been shown that no criterion is able to predict the location of nucleation, the crack angle and load leading to crack initiation together.

In this study we consider the SWT parameter despite the fact that it has been shown to be unable to predict the crack initiation angle (the investigated cracks are very often inclined toward the centre of the contact). The hypothesis to identify the crack initiation plane as the critical plane of the criterion is not obvious and the crack initiation process may be much more complicated as recently pointed out by another study \cite{Proudhon2007}. On the other hand the SWT criterion has a low cycle fatigue behaviour which has been shown to be effective for finite life \cite{Fouvry2002} and also was used recently be Fridrici et al. to tackle the same kind of problems in titanium alloys \cite{Fridrici2005}. In addition, the material coefficients for this criterion are given by Szolwinski et al. \cite{Szolwinski1996}.

\subsection{Prediction of the crack nucleation boundary at 50000 fretting cycles}
\label{50000.ssc}

The use of the SWT criterion for fretting is now well documented (see for example \cite{Sum2005,Proudhon2005,Fridrici2005}). Thus, only the general equation is reproduced here. According to this model, the initiation is likely to occur on the plane where the SWT parameter is maximum. The SWT parameter $\Gamma$ is evaluated in the current plane by the product between the amplitude of the strain $\varepsilon _a$ and the maximum stress normal to this plane $\sigma_{max}$.

\begin{equation}
\Gamma=\sigma_{max}\times\varepsilon_a=\frac{(\sigma _f')^2}{E}(2N)^{2b'}+\sigma _f'\varepsilon _f'(2N)^{b'+c'}
\label{SWT.eqn}
\end{equation}

where $\sigma_f'$ is the fatigue strength coefficient, $b'$ is the fatigue strain exponent, $\varepsilon_f'$ is the fatigue ductility coefficient and $c'$ is the fatigue ductility exponent. The mechanical and fatigue properties of the studied alloy are listed in tables \ref{alu_mecha.tab} and \ref{alu_fatigue.tab} respectively.

\linespread{1}
\begin{table}[!htb]
\caption{Fatigue properties of the studied Al2024 (from \cite{Szolwinski1996})}
\begin{tabular}{cccc}
\hline
$\sigma_f'$ (MPa) & $b'$ & $\varepsilon_f'$ & $c'$ \\
\hline
714 & -0.078 & 0.166 & -0.538 \\
\hline
\end{tabular}
\label{alu_fatigue.tab}
\end{table}

This approach has been previously used to study the crack nucleation at 50\,000 cycles \cite{Proudhon2005}. The results can be summarized as follow:

\begin{itemize}
\item the crack nucleation threshold predicted by the analysis of the local analytical stresses is very far from the experimental result ($\simeq100$ N/mm to be compared to $Q_{th}=240$ N/mm).
\item performing size effect calculations allowed for fitting of the average experimental result but the precise behaviour (i.e. the effect of the loading pressure) cannot be captured.
\item the average process volume showing the best correlation was found to be correlated to the mean grain radius (which was measured by Electron BackScattered Diffraction analysis giving 75~\textmu m).
\end{itemize}

The physical meaning of the process volume size is still an open question. The most often quoted significance is the grain size. Indeed it is argued here that the stress/strain state inducing initiation must be sufficiently widespread to make the very short crack propagate in the adjacent grain. Looking at several studies on different materials, a large range of sizes have been used for the process volume. 5~\textmu m in steel \cite{Fouvry2002}, 30~\textmu m in titanium alloy \cite{Fridrici2005} and 80~\textmu m in aluminium alloy \cite{Proudhon2005}, each time correlated to the grain size. On the other hand, the grain size cannot be the only significant parameter due to other cases where no correlation has been found. Moreover, this parameter does not take into account any mechanical quantity such as for example the slip amplitude which is certainly relevant for the crack initiation condition.

\begin{figure}[!hbt]
\centering
\includegraphics[width=75mm]{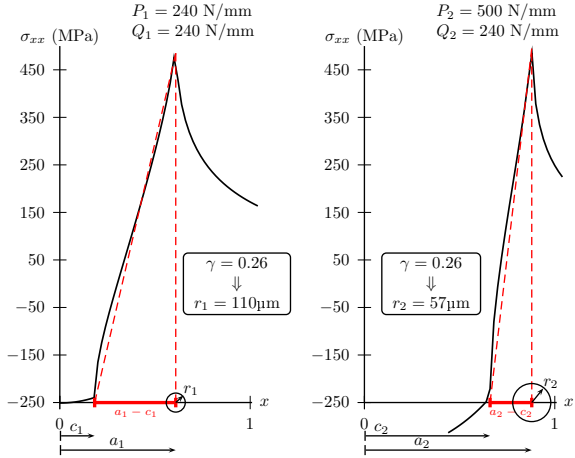}
\caption{Illustration of the radius of the process volume needed to average the stress gradient with two different normal loads $P_1$ and $P_2$.}
\label{s11_plot_P1P2.fig}
\end{figure}

We consider here a new approach using a variable process volume. This idea comes from the fact that the crack initiation may be strongly monitored by the stress state in the slip zone only. More precisely, the severity of the stress gradient located in the slip zone may be responsible of the crack initiation risk. This is illustrated by figure~\ref{s11_plot_P1P2.fig} where two distributions of the surface traction\footnote{One should note that the $\sigma_{xx}$ component is dominant in the slip zone, compared to the other stress values, (they vanish when $x\mapsto a$ where the stress state becomes purely uniaxial)} ($\sigma_{xx}$) are plotted for the same conditions but with different normal loads ($P_1<P_2$).

\begin{figure}[!hbt]
\centering
\includegraphics[]{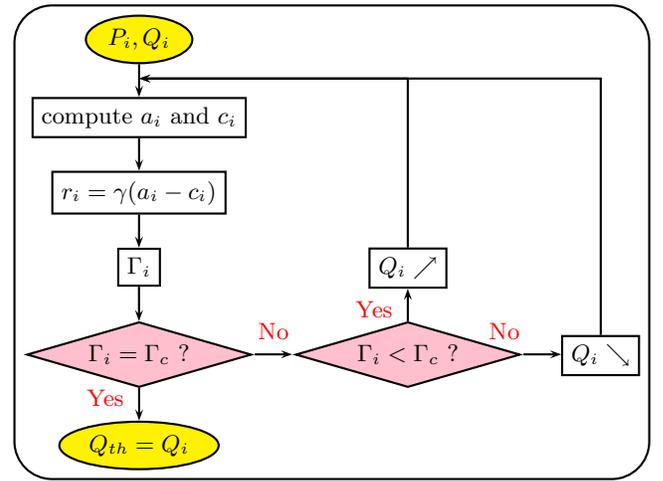}
\caption{Flowchart of the different steps required to predict the critical load $Q_{th}$ for each pressure level $P_i$~; the third step has been added to the conventional constant process volume approach.}
\label{flowchart.fig}
\end{figure}

\begin{figure}[!hbt]
\centering
\includegraphics[]{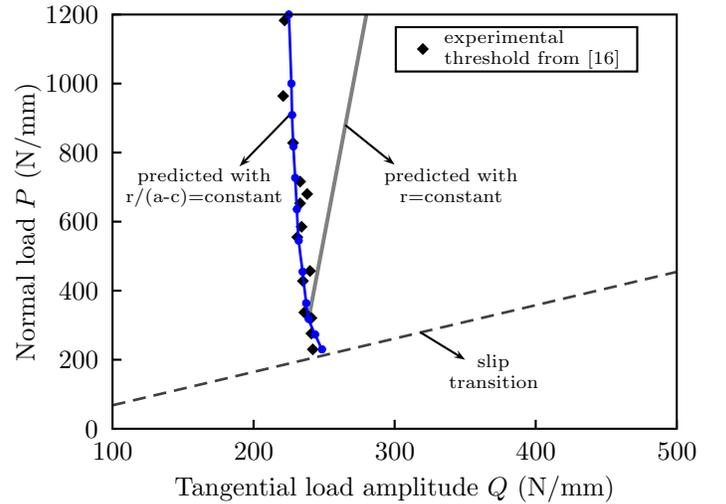}
\caption{Prediction of the crack nucleation boundary at 50\,000 fretting cycles with the variable process volume approach.}
\label{prediction_r_variable.fig}
\end{figure}

It is clear that the severity of the stress gradient can be related to the slip zone width $a-c$ due to the quasi-linear evolution of $\sigma_{xx}$. Keeping the process volume size constant would introduce a strong effect of the pressure, which can actually be seen looking at the solid line prediction replotted on fig.~\ref{prediction_r_variable.fig} for the calculation with $r=80$ \textmu m. From here the radius of the process volume zone is no longer constant. We introduce a new variable $\gamma$ defined as the ratio of the process volume radius $r$ and the slip zone width $a-c$:

\begin{equation}
\gamma=\frac{r}{a-c}
\end{equation}
The value of $\gamma$ is identified once for all for the conditions which gave r=80\textmu m. For the corresponding loading conditions ($P=318$ N/mm, $Q=240$ N/mm) one can calculate $a=700$ \textmu m and $c=393$ \textmu m. This leads to the value of $\gamma=0.26$. The crack nucleation threshold is then computed through this new approach according to the flowchart shown in fig.~\ref{flowchart.fig}.
\par Prediction of the crack nucleation boundary for 50\,000 fretting cycles with $\gamma=0.26$ is presented in fig.~\ref{prediction_r_variable.fig}. This approach gives, a very good correlation with the experimental results. In particular, the pressure effect is well described.
\par This result highlights the reliability of the approach and shows that the process volume size, in addition to be related to some microstructural characteristic of the material, must be linked to the mechanical state as well, such as the slip zone width, to be able to predict the crack nucleation precisely.
\subsection{Prediction of the low cycle fretting behaviour}
\par
\begin{figure}[!hbt]
\centering
\includegraphics[width=75mm]{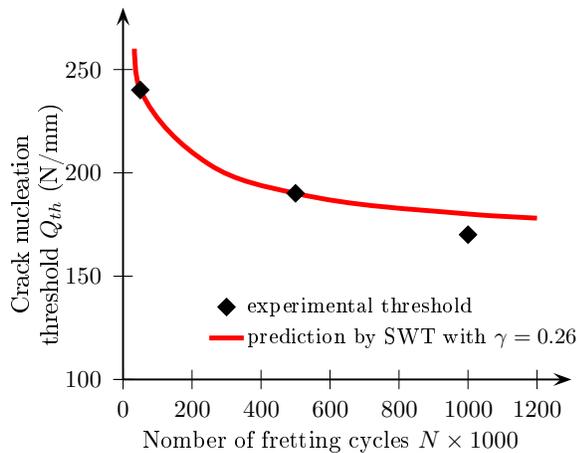}
\caption{Prediction of the evolution of $Q_{th}$ with the number of cycles, by application of the SWT criterion suited with a variable process volume size.}
\label{fretting_SWT_vs_N.fig}
\end{figure}

 The same model is applied to test the experimental results obtained for $5.10^5$ and $10^6$ fretting cycles (see \S \ref{exp_N.sec}). The $\gamma$ parameter is kept constant at the same value identified for 50\,000 cycles ($\gamma=0.26$, see \S \ref{50000.ssc}). The prediction is done in the same way by varying the number of cycles in equation \ref{SWT.eqn}. The results of these computations are gathered in figure~\ref{fretting_SWT_vs_N.fig}.
\par The experimental behaviour appears well correlated with the prediction of the SWT parameter. In particular one can see a very rapid decrease of the critical tangential force needed to nucleate a fretting crack, as the number of cycles increases. Around $10^6$ cycles a plateau is reached, corresponding to endurance conditions. This behaviour is consistent with another study on the same alloy and another 7xxx series alloy \cite{Munoz2006} where $10^6$ fretting cycles are clearly identified as a stabilized condition.
\section{Conclusion}
\par Two main results have been presented in this paper. On the experimental point of view, a new representation of the crack nucleation condition is introduced through the $(P,Q,N)$ diagram, leading to a concrete and complete description of the material fretting resistance to initiation and this with a limited, although still quite important, number of tests. In order to predict this initiation condition, the SWT criterion is used and the classical computation is extended with a variable process volume size. This further opens the question of the significance of this parameter as it appears not to be only related to a microstructure characteristic length. The use of the slip zone width to determine the process volume radius clearly identifies a \emph{mechanical} significance. Eventually the final answer may require a mix of the microstructural/mechanical behaviour. This could at last be answered by extending the approach to different materials as for instance steels and titanium alloys.
\bibliography{fretting_swt}

\end{document}